# Directly stabilized solitons in silicon-nitride microresonators

**Chengying Bao,[1*] Yi Xuan,[1,2] Daniel E. Leaird,[1] Minghao Qi,[1,2] and Andrew M. Weiner[1,2]**

[1]*School of Electrical and Computer Engineering, Purdue University, 465 Northwestern Avenue, West Lafayette, Indiana 47907-2035, USA*
[2]*Birck Nanotechnology Center, Purdue University, 1205 West State Street, West Lafayette, Indiana 47907, USA*
*\*bao33@purdue.edu*

**Abstract:** We investigate soliton generation dynamics with the influence of thermal effects. Either soliton annihilation or survival can occur in different trials with the same tuning method, and a spontaneous route to soliton formation is observed.

## 1. Introduction

Soliton generation in microresonators has been an important way to achieve coherent Kerr combs with broad-bandwidth. When solitons are formed from the chaotic state, there is a significant drop in the intracavity power, which can blue-shift the cavity resonance and cause transient instability of soliton generation in microresonators with strong thermal effects such as silicon-nitride (SiN) cavities [1]. To overcome this transient instability, methods such as "power-kick" [1] and backward tuning [2] have been developed. The "power-kick" method needs additional devices (acoustic-optical modulators or electro-optical modulators), while the backward tuning is limited by the response time of the laser and the duration of the soliton-step. For compact integration of soliton Kerr combs, it is desired to avoid additional devices and have a straightforward way for soliton generation. Therefore, investigation of thermal effects in soliton generation dynamics is important. Here, we report on direct soliton generation (DSG, defined as tuning the laser and stopping at a specific wavelength for generation of one or multiple solitons) in SiN microresonators. We find soliton annihilation and survival can occur in different trials of the same scan method, and spontaneous soliton generation is possible.

## 2. Simulations of direct soliton generation

Thermal effects are important for the soliton generation. To unveil its role in DSG, we include thermal effects in numerical simulations based on the Lugiato-Lefever equation (LLE) [2, 3] as,

$$\left(t_R \frac{\partial}{\partial t} + \frac{\alpha+\kappa}{2} + i\frac{\beta_2 L}{2}\frac{\partial^2}{\partial \tau^2} + i(\delta_0+\delta_\Theta)\right)E - i\gamma\left(1+\frac{i}{\omega_0}\right)L\left(E\int_{-\infty}^{+\infty} R(\tau')|E(t,\tau-\tau')|^2 d\tau'\right) - \sqrt{\kappa P_{in}} = 0, \qquad (1)$$

$$\frac{d\delta_\Theta}{dt} = -\frac{\delta_\Theta}{\tau_0} + \xi P. \qquad (2)$$

Here, $E$, $t_R$, $\alpha$, $\kappa$, $\beta_2$, $\gamma$, $L$, $\omega_0$, $P_{in}$, are the envelope of the intracavity field, round-trip time, intrinsic loss, coupling coefficient, dispersion, nonlinear coefficient, cavity length, pump frequency, coupled power, respectively. $R(\tau)$ is the nonlinear response, including the instantaneous electronics response and delayed Raman response, which is calculated in the same way as in [4]. $\delta_0$ is the pump laser phase detuning, $\delta_\Theta$ is the thermal phase detuning. The thermal phase detuning dynamics is included in eq. 2, where $\tau_0$ is the thermal response time, $\xi$ is a coefficient describing how average intracavity power ($P$) will affect the phase detuning. We choose $\alpha+\kappa=0.0037$, $\kappa P_{in}=0.11$ mW, $\beta_2=-61$ps$^2$/km, $\gamma=1.4$ (Wm)$^{-1}$, $L=628$ μm, $\tau_0=100$ ns, $\xi=-4.5*10^3$ (Ws)$^{-1}$ and start comb generation from noise by tuning $2\delta_0/(\alpha+\kappa)$ from -3 to 25 in 2 μs and holding $2\delta_0/(\alpha+\kappa)$ at 25 for another 2 μs for the simulation of soliton generation dynamics.

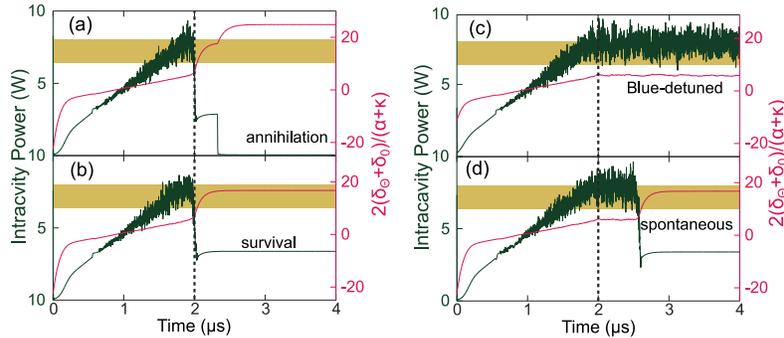

Fig. 1 Soliton generation dynamics in simulations, (a) soliton annihilation, (b) soliton survival, (c) comb remains in the chaotic regime (d) spontaneous soliton generation after laser scan ends. The dashed line shows the end of laser scan and the shaded boxes show the detuning ranges where the soliton can be supported. The olive lines are the intracavity power change and the magenta lines are the detuning change.

Solitons can only be supported in a narrow range of detuning (referred as referred as soliton supporting range, SSR). The SSR is the mapped to be $9.2<2(\delta_0+\delta_\Theta)/(\alpha+\kappa)<17.9$, with thermal effects turned off. Depending on the relationship between the $2(\delta_0+\delta_\Theta)/\alpha$ and SSR, various soliton dynamics are found in DSG. In some cases, the thermally induced resonance blue-shift associated with soliton formation can cause $2(\delta_0+\delta_\Theta)/(\alpha+\kappa)$ to exceed the SSR and the solitons annihilate (Fig. 1(a)); in others $2(\delta_0+\delta_\Theta)/(\alpha+\kappa)$ can remain in the SSR and solitons survive in DSG (Fig. 1(b)). Moreover, when the laser scan ends with the comb in the chaotic regime, in some trials the comb remains in the chaotic regime and solitons do not form (Fig. 1(c)), whereas in some trials the comb evolves to the soliton state spontaneously after the laser scan ends (Fig. 1(d)). Simulation results show the strength of the thermal effects, $\xi$, is important for the stability of solitons in DSG. Weak thermal effects usually facilitate soliton survival while strong thermal effects tend to lead to soliton annihilation. This results show the importance of reducing the material loss converted to heat in cavity fabrication.

## 3. Experimental demonstration of direct soliton generation

In experiments, we generate the comb in a SiN microresonator with loaded Q-factor of 1 million and geometry of 800*2000 nm. We tune the laser from blue to red for direct soliton generation. The cavity generally supports multi-soliton initially in DSG with the spectrum shown in Fig.2(a). After multi-solitons are stabilized, a single soliton can be realized by backward tuning of the laser (Fig. 2(b)).

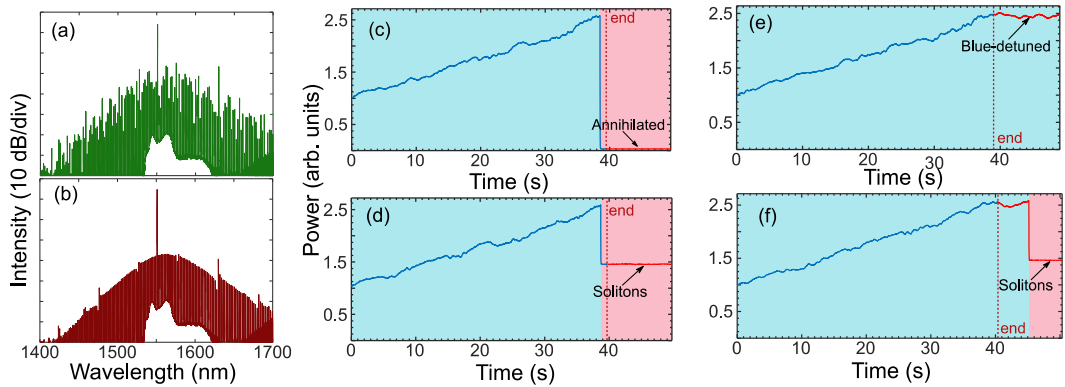

Fig. 2 Experimental results of DSG. Spectrum of (a) multi-soliton (b) single-soliton. (c) Soliton annihilation and soliton forms before laser scan ends. (d) Soliton survival with soliton formation before laser scan ends. (e) Comb remains in the chaotic regime. (f) Comb remains chaotic when laser scan ends and forms soliton spontaneously. The dashed lines in (c-f) indicate end of the laser scan and blue lines power trace recorded before laser scan ends and red lines are recorded after laser scan ends. The blue-shaded boxes show blue-detuning with respect to the hot resonance (the resonance as affected by both Kerr and thermal effects), while the red-shaded boxes are for red-detuning.

Various routes to soliton generation are also observed in a series of experiments with the same laser tuning protocol. Solitons can be formed before the laser scan ends, as shown in Figs. 2(c, d). Similar to simulations, solitons may either annihilate subsequently (Fig. 2(c)) or survive (Fig. 2(d)). Hence either soliton annihilation or survival are observed under the same experimental conditions. Also, the comb can remain in the chaotic regime when the scan ends, as shown in Figs. 2(e, f). When the comb is chaotic when laser scan ends, solitons can be formed spontaneously, as predicted in simulations. The experimental results verify the simulation results and help unveil the role of thermal effects in DSG.

## 4. Conclusion

We have shown solitons can be directly stabilized in SiN microresonators with a simple experimental method, despite the large thermo-optical coefficient of SiN. We show that either soliton survival or annihilation can occur when tuning the laser in the same method in DSG. In addition, we find a spontaneous route to soliton formation. Our results give important insights into soliton generation in microresonators, which can contribute to realizing compact integration of single or multiple soliton Kerr combs.